\documentclass[12pt]{article}

\usepackage{amssymb}
\usepackage{bbm}
\usepackage{epsfig}
\usepackage{dsfont}
\usepackage{amstext}
\usepackage{amsmath}
\usepackage{psfrag}
\usepackage{a4}
\usepackage{a4wide}
\usepackage{feynmp-auto}

\usepackage{amsmath, amssymb, graphics, setspace}
\usepackage{amsfonts}
\usepackage{multirow}
\usepackage{slashed}
\usepackage{amssymb}
\usepackage{array}
\usepackage{ragged2e}
\usepackage{float}
\usepackage{graphicx}
\graphicspath{ {Images/} }
\graphicspath{ {feyn/} }
\usepackage{xparse}
\usepackage{amsmath, amssymb, graphics, setspace}
\usepackage{rotating}
\usepackage{float}
\usepackage{graphicx}
\usepackage{graphics}
\usepackage{makeidx}
\RequirePackage{luatex85}
\usepackage{tikz}
\usepackage{hyperref}
\usepackage{subfig}
\usepackage[utf8]{inputenc}
\newcommand{\mathsym}[1]{{}}
\newcommand{\unicode}[1]{{}}

\title{}
\begin{document}
 	\begin{center}
		\baselineskip 20pt 
		{\Large\bf Pseudosmooth Tribrid Inflation in $SU(5)$}
		
		\vspace{1cm}
		
		{\large 
			Muhammad Atif Masoud$^{a,}$\footnote{E-Mail: atifmasood23@gmail.com}, Mansoor Ur Rehman$^{a,}$\footnote{ E-mail: mansoor@qau.edu.pk}
			and Qaisar Shafi$^{b,}$\footnote{ E-mail: shafi@bartol.udel.edu} 
		} 
		\vspace{.5cm}

		{\baselineskip 20pt \it
			$^a$Department of Physics, Quaid-i-Azam University , \\ 
			Islamabad 45320, Pakistan \\
			\vspace{2mm} 
			$^b$Bartol Research Institute, Department of Physics and Astronomy, \\
			University of Delaware, Newark, DE 19716, USA \\
		}
		
		\vspace{1cm}
	\end{center}
	
\begin{abstract}
A realistic tribrid model of sneutrino inflation is constructed in an $R$-symmetric $SU(5)$ grand unified theory (GUT). To avoid the monopole problem, a pseudosmooth inflationary trajectory is generated with the help of an additional $Z_5$ symmetry which is broken during and after inflation. The predictions of inflationary parameters are made at the central value of the scalar spectral index, $n_s =0.968$. The largest possible value of the tensor to scalar ratio, $r\lesssim 0.0027$, is obtained with sub-Planckian field values ($\lesssim m_P$). A successful realization of reheating and leptogenesis is achieved by avoiding the gravitino problem with a reheat temperature as low as $10^6$ GeV. The predicted range of the gauge symmetry breaking scale, $ 5 \times 10^{16} \lesssim M/\text{GeV} \lesssim  5 \times 10^{17}$, turns out to be somewhat larger than the typical GUT scale. With additional vector-like families, a successful gauge coupling unification is achieved by avoiding the no-go theorem related to $R$-symmetric $SU(5)$ GUT. 
\end{abstract}	

\section{Introduction}
An interesting extension of supersymmetric hybrid inflation \cite{Dvali:1994ms,Copeland:1994vg,Rehman:2009nq,Linde:1997sj,Senoguz:2004vu} is tribrid inflation \cite{Antusch:2004hd,Antusch:2008pn,Antusch:2009ef,Antusch:2012bp,Antusch:2009vg} where a matter field can be employed to realize inflation. One of the simplest candidates for tribrid (matter) inflation could be a sneutrino, the superpartner of the right handed neutrino. An early model of sneutrino inflation was proposed in \cite{Murayama:1992ua} as a chaotic model of inflation. [Also see \cite{Ellis:2003sq} where the various predictions of this model were compared with the available experimental data.] This model, however, is plagued with the common problems of realizing chaotic inflation in a supergravity framework \cite{Lyth:1998xn}. The first model of sneutrino tribrid inflation was introduced by in \cite{Antusch:2004hd}. This framework, however, is not suited for realizing inflation in a grand unified theory (GUT) model since the gauge symmetry associated with the waterfall GUT Higgs field breaks down at the end of inflation, and so the monopole problem is not resolved. In addition, a domain wall problem arises from the spontaneous breaking of a $Z_4$ symmetry which is introduced to constrain the structure of the superpotential. In order to resolve this problem higher order $Z_4$ symmetry breaking terms are introduced. For a general discussion of tribrid inflation see \cite{Antusch:2012bp,Antusch:2012jc}, where three types of tribrid inflation are identified, depending on terms of different origin dominating the scalar potential. From these scenarios, only pseudosmooth tribrid inflation \cite{Antusch:2012bp} is well suited for GUTs with the potential monopole problem. In pseudosmooth tribrid inflation, a shifted smooth track is employed for inflation with the GUT symmetry broken during inflation, such that the monopoles produced during inflation are inflated away. 

In this paper we study the possibility of realizing sneutrino tribrid inflation in $SU(5)$ GUT. A pseudosmooth tribrid inflation model employing a $Z_5$ symmetry is particularly suited for the $SU(5)$ case. A non-minimal K\"{a}hler potential is required for the realization of this model. Including a supergravity mass term for the waterfall GUT Higgs field a shifted smooth track, suitable for inflation, can be generated. On this track the $SU(5)\times Z_5$ symmetry is broken and, therefore, any defects produced during inflation are inflated away. Another common problem in an $R$-symmetric $SU(5)$ GUT is the presence of light triplet and octet fields \cite{Khalil:2010cp}, so that a successful gauge coupling unification in  minimal supersymmetric standard model (MSSM) is spoiled. According to a no-go theorem discussed in \cite{Barr:2005xya,Fallbacher:2011xg}, this is a generic problem of $R$-symmetric GUTs based on a simple group. This problem is, however, circumvented in our model with the help of additional vector-like families and as we shall show, a successful gauge coupling unification is achieved. Moreover, assuming relatively large squarks/sleptons masses of order $10$ TeV or so, the dimension five proton decay rate is suppressed in accordance with the experimental bound \cite{Miura:2016krn}.

An attractive feature of sneutrino tribrid inflation is the realization of reheat temperature as low as $10^6$ GeV. This feature naturally avoids the gravitino problem usually encountered in supergravity models of inflation. A model of non-thermal leptogenesis \cite{Lazarides:1991wu} is employed in order to explain the observed baryon asymmetry. The numerical predictions of the various inflationary observables are found to be perfect agreement with the latest Planck 2018 results \cite{Aghanim:2018eyx,Akrami:2018odb}. In particular, a tensor to scalar ratio $r \approx 0.0027$ can be obtained, and this hopefully can be tested in future experiments \cite{Andre:2013afa,Matsumura:2013aja}.

\section{Superpotential for Tribrid Inflation in $SU(5)\times Z_5$ Model}
	   The minimal supersymmetric standard model (MSSM) matter content with right handed neutrinos are embedded into $\bar5_{i}$, $10_{i}$ and $1_{i}$ dimensional representations of supersymmetric SU(5) as 
	\begin{eqnarray}\label{eq:1}
	\bar5_{i}&=&D_{i}^{c}(\bar3,1,1/3)+L_{i}(1,2,-1/2),\notag\\
	10_{i}&=&Q_{i}(3,2,1/6)+U_{i}^{c}(\bar3,1,-2/3)+E_{i}^{c}(1,1,1),\notag\\
	1_{i}&=&N_{i}=\nu_{i}^{c}(1,1,0),
	\end{eqnarray}
where $i$ is the generation index $(i = 1,2,3)$ and $N_i=\nu_{i}^{c}$ represents the right handed neutrino superfield. The GUT Higgs superfield, $ 24_{H}$, is responsible for the breaking of $SU(5)$ into MSSM  whereas the electroweak Higgs doublets ($H_u$, $H_d$) contained in the $5_{H}$ and $\bar5_{H}$  Higgs superfields trigger the electroweak breaking. The decomposition of minimal Higgs sector in terms of MSSM superfields is given by
 \begin{eqnarray}
5_H &=& {H_T(3,1,-1/3)} + H_u (1,2,1/2),  \nonumber \\
\overline{5}_H &=&  \overline{H}_T(\bar{3},1,1/3) + H_d(1,2,-1/2), \nonumber \\
24_H &=& H_{24}(1,1,0) + W_H(1,3,0) + G_H(8,1,0) + X_H(3,2,-5/6) + \overline{X}_H(\overline{3},2,5/6).
 \end{eqnarray}

The desired superpotential of an $R$-symmetric $SU(5)\times Z_5$ model, including a gauge-singlet superfield $S$, can be written as
\begin{eqnarray}\label{eq2}
W&=&\kappa S\left(\mu^2 + \frac{Tr(24_{H}^{5})}{m_P{}^3} + \alpha\frac{Tr(24_{H}^{2})Tr(24_{H}^{3})}{m_P{}^3}\right) - \beta_{ij}\frac{Tr(24_{H}^{3})}{m_P{}^2}N_{i}N_{j}\nonumber \\
&+&\lambda_{1} \frac{Tr(24_{H}^{2})}{m_{P}}\overline 5_{H}5_{H}+\frac{\lambda_{2}}{m_{P}}\overline 5_{H}24_{H}^{2}5_{H} \nonumber  \\
&+& y_{ij}^{(u)}10_{i}10_{j}5_{H}+y_{ij}^{(d,e)}10_{i}\overline 5_{j}\overline 5_{H}+\frac{\lambda_{ij}^{\nu}}{m_{P}^{2}} Tr(24_{H}^{2}) N_{i}\overline 5_{j}5_{H}
+\frac{\tilde\lambda_{ij}^{\nu}}{m_{P}^{2}}N_{i}\overline 5_{j}24_{H}^{2}5_{H},
\end{eqnarray}
where $\mu$ is a superheavy mass, $m_{P} = 2.43\times 10^{18}$ GeV is the reduced Planck mass and all other couplings ($\kappa, \alpha, \beta_{ij}, \lambda_1,\lambda_2,\cdots$) are dimensionless. The charge assignments of the various superfields under $U(1)_{R}$ and $Z_{5}$ symmetries are respectively given by
\begin{eqnarray}\label{eq:3}
R(S,24_{H},5_{H},\bar 5_{H},10_{i},\bar 5_{i},N_{i}) &=& \left(1,0,\frac{2}{5},\frac{3}{5},\frac{3}{10},\frac{1}{10},\frac{1}{2}\right), \nonumber\\
q_{5}(S,24_{H},N_{i},5_{H},\bar 5_{H},10_{i},\bar 5_{i}) &=& (0,1,1,3,0,1,4),
\end{eqnarray}
with $R(W)=1$. The terms in the first line of the superpotential $W$ are relevant for tribrid inflation which is discussed below in detail. Owing to $SU(5)$ gauge invariance of the superpotential it is required to align Higgs $24_H$ superfield along the standard model (SM) gauge singlet direction, $H_{24}$, as
   \begin{eqnarray}\label{eq:4}
24_{H}& \longrightarrow & H_{24} = \frac{h}{\sqrt{15}}(1,1,1,-3/2,-3/2).
\end{eqnarray}
The global supersymmetric minimum, therefore, occurs at
\begin{eqnarray}\label{min}
\left< h^{5} \right> \equiv M^5 = \frac{8\sqrt{15}}{\left(\frac{13}{30} + \alpha\right)} \mu ^2 m_{P}^{3},\qquad \left< S \right> = 0,\qquad\left< N_{i} \right> = 0,
\end{eqnarray}
for the relevant superfields. The importance of the various terms in the superpotential can now be described conveniently in terms of $h$ and its vacuum expectation value $M$. 

The terms in the second line of Eq.~(\ref{eq2}),
\begin{equation}
W \supset \frac{h^2}{m_P} \left(  \left(\frac{\lambda_{1}}{2}+\frac{\lambda_{2}}{15} \right) \overline H_{T} H_{T} + \left(\frac{\lambda_{1}}{2}+\frac{3\lambda_{2}}{20} \right) H_{u} H_{d} \right) \supset
\mu_2 H_{u} H_{d} + \mu_3 \overline H_{T} H_{T} ,
\end{equation}
are relevant for the doublet-triplet problem. Here, the mass parameter, $\mu_2$, is just the $\mu$-parameter of MSSM which is usually taken to be of electroweak scale with $\lambda_1 \simeq -3\lambda_2/10$. On the other hand,  the mass parameter, $\mu_3 \simeq -\lambda_1 (M/m_P) M /12$, is  taken to be order GUT scale in order to suppress dimension-5 proton decay amplitude mediated by the color triplet Higgs pair. This further requires the squark/slepton masses to be $\gtrsim 10$ TeV. Therefore, the doublet-triplet problem is solved, as usual, by fine tuning. 
Lastly, the couplings, $y_{ij}^{(u)},y_{ij}^{(d,e)},\lambda_{ij}^{(\nu)},\tilde\lambda_{ij}^{(\nu)}$, in the third line of Eq.~(\ref{eq2}) include the quark and lepton Yukawa couplings. In order to obtain the observed tiny neutrino masses, Majorana mass terms for the right handed neutrinos are required. Even though an explicit Majorana mass term is not allowed due to $Z_5$ symmetry, the spontaneous breaking of $SU(5)$ gauge symmetry generates an effective Majorana mass term, $(1/2) M^R_{ij} N_i N_j$, with
   \begin{eqnarray}\label{eq:6}
M^R_{ij} & = &\frac{\beta_{ij}}{2\sqrt{15}} \left( \frac{M}{m_P} \right)^2 M,
\end{eqnarray}
from the last term in the first line of Eq.~(\ref{eq2}). Taking Majorana masses to be order $10^{13}$ GeV the light neutrino masses are naturally explained via type-I seesaw mechanism. As we discuss below, this term also plays an important role in realizing sneutrino inflation and subsequent reheating.

\section{Inflationary Scalar Potential}
To discuss inflation we consider the following superpotential terms from Eq.~(\ref{eq2}),
\begin{eqnarray}\label{eq:7}
W &\supset& \kappa S\left(\mu^2 + \frac{Tr(24_{h}^{5})}{m_P{}^3} + \alpha\frac{Tr(24_{h}^{2})Tr(24_{h}^{3})}{m_P{}^3}\right) - \beta_{ij} \frac{Tr(24_{h}^{3})}{m_P{}^2}N_i N_j,\nonumber\\
&\supset& \mu^2 S\left(1-\left(\frac{h}{M}\right)^{5}\right)+\beta (\mu^2 m_P) \left( \frac{h}{M} \right)^3 \left( \frac{N}{M} \right)^2,
\end{eqnarray}
where $\beta = \frac{2\beta_{11}}{\left(\frac{13}{30} + \alpha\right)} $, $N \equiv N_1$ and to achieve $M \ll m_P$ with a natural value of $\kappa$ we set $\kappa = 1$. With smaller values of $\kappa$, the value of $M$ becomes Planckian. As there is no contribution from the relevant fields in the $D$-term scalar potential, the global SUSY scalar potential obtained from the $F$-term is given by, 
\begin{equation}
V_F = \mu^4 \Bigg(\left|  1-\left(\frac{h}{M}\right)^5\right|^2+  \left| 3 \beta \left(\frac{m_{P}}{M}\right)\left(\frac{ N^2 h^2 }{M^4}\right)- \frac{5Sh^4}{M^5}\right|^2 +\left| 2 \beta \left(\frac{m_P}{M}\right) \left(\frac{Nh^3 }{M^4}\right)\right|^2
 \Bigg),
\end{equation}
where, $V_F = |\partial W/ \partial z_i|^2$, with $z_{i} \in ( S,h,N )$. To keep the discussion simple we assume that the phases of the fields have been stabilized before the start of observable inflation and, therefore, the above potential reduces to the following form,
\begin{equation}
V_F = \mu^4 \Bigg(\left( 1 - z^5 \right)^2+  \left( 3 \beta \left(\frac{m_{P}}{M}\right) y^2 z^2 - 5 x z^4 \right)^2 +\left( 2 \beta \left(\frac{m_P}{M}\right) y z^3 \right)^2
 \Bigg),
\end{equation}
where, 
\begin{eqnarray}\label{eq:9}
x =\frac{|S|}{M},\quad y=\frac{|N|}{M},\quad z=\frac{|h|}{M}.
\end{eqnarray}

Next we aim to find an effective single field form of the above potential, and to achieve this goal we need to include supergravity (SUGRA) corrections which are obtained from the following formula,
\begin{equation}
V_F = e^{K/m_{P}^{2}}\left(K_{ij}^{-1}D_{z_{i}}WD_{z_{j}^{*}}W^{*} -3m_{P}^{-2}|W|^{2}\right),
\end{equation}
where
\begin{eqnarray}
D_{z_{i}}W = \frac{\partial W}{\partial z_{i}}+\frac{1}{m_{P}^{2}}\frac{\partial K}{\partial z_{i}}W,
\quad K_{ij} = \frac{\partial^{2}K}{\partial z_{i}\partial z_{j}^{*}}, \quad
D_{z_{j}^{*}}W^{*} = (D_{z_{i}}W)^{*}.
\end{eqnarray}
Here, we consider the following power-law expansion of the K\"{a}hler potential
\begin{eqnarray}\label{kahler}
K & = & |S|^{2}+|N|^{2}+Tr|24_H|^{2}\nonumber\\
&+&\kappa_S \frac{|S|^{4}}{4m_{P}^{2}}+\kappa_{N} \frac{|N|^{4}}{4 m_{P}^{2}} +\kappa_{h} \frac{(Tr|24_H|^{2})^{2}}{4 m_{P}^{2}} \nonumber\\
&+& \kappa_{SN} \frac{|S|^{2}|N|^{2}}{m_{P}^{2}}+\kappa_{S h} \frac{|S|^{2}Tr|24_H|^{2}}{m_{P}^{2}}+\kappa_{N h} \frac{|N|^{2}Tr|24_H|^{2}}{m_{P}^{2}} \cdots. 
\end{eqnarray}
Including only the relevant SUGRA correction terms, the scalar potential as a function of the three fields is given by
\begin{eqnarray}\label{eq:12}
V_{3}(x,y,z) & =& \mu^{4} \Bigg( \left(1-  z^5 \right)^{2}+ \left (  3\beta\left(\frac{m_{P}}{M}\right)y^2 z^2  - 5xz^4 \right)^2  + \left (2 \beta \left(\frac{m_P}{M}\right) yz^3 \right)^2 \nonumber \\
& - &   \kappa_S \left(\frac{M}{m_P}\right)^{2} x^2 + \kappa_{h} \left(\frac{M}{m_P}\right)^{2} z^2 +\gamma\left(\frac{M}{m_P}\right)^{2} y^2 +\delta \left(\frac{M}{m_P}\right)^{4} y^4+\cdots \Bigg),  \qquad
\end{eqnarray}
where $\gamma=1-\kappa_{SN}$ and $\delta=\frac{1}{2}+\kappa_{SN}^{2}-\kappa_{SN}+\frac{1}{4}\kappa_{N}$. 

\subsection*{Stabilization of S ($x = S/M$) Field}
In order to obtain an effective single-field potential we first minimize the three-field potential $V_3$ with respect to $x$. The potential minimum occurs at,
\begin{align}\label{eq:14}
&x_{min } = \frac{15  y^2 z^6 \beta m_{P}^3}{25 m_P^2 z^8-M^2 \kappa _{S}},
\end{align}
with $\beta > 0$ and $\kappa_S <0$. The mass squared of the $S$ field, $m_S^2$, in term of Hubble mass squared, $H^{2} \simeq \frac{\mu ^4}{3 m_P^2}$, is given by
\begin{eqnarray}\label{eq:15}
m_{S}^{2}/H^{2} \simeq \left(75\left(\frac{m_P}{M}\right)^{2}z^8-3\kappa _S \right).
\end{eqnarray}
Therefore, the $S$ field attains Hubble size mass for $\kappa _S \lesssim-\frac{1}{3}$ and quickly settles down to its minimum. This leads us to the following effective two-field potential,
\begin{figure}[t]
	\centering
	\includegraphics[width=10cm]{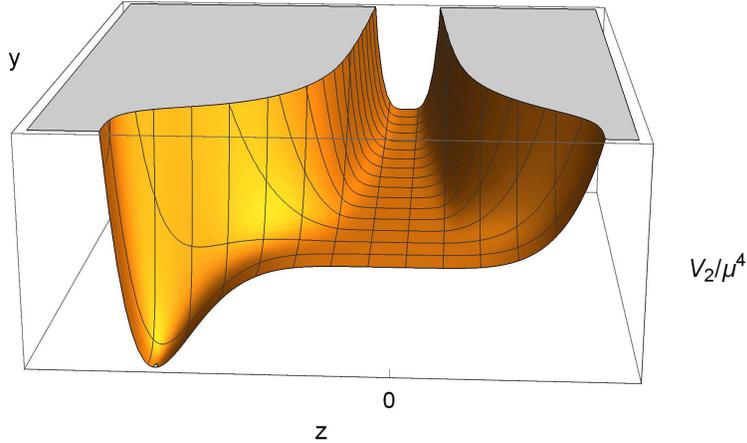}
	\caption{The normalized two-field potential $V_{2}/\mu^4$ as a function of $z = \frac{h}{M}$ and $y = \frac{N}{M}$ with $x=\frac{S}{M}=x_{min}$, $\kappa_h = -1$, $\kappa_S = -1/3$, $\gamma = \delta = 0$ and $M = 1\times10 ^{17}$ GeV. The zoom-in plot of pseudosmooth inflationary valley with $z \neq 0$ is shown in Fig.~\ref{Fig2}.}   
	\label{Fig1}
\end{figure}
\begin{eqnarray}\label{eq:16}
V_{2}(y,z)&\equiv& V_{3}(x_{\min },y,z)   \nonumber \\
   &=&  \mu^{4}\Bigg( \left(1-  z^5 \right)^{2}+ \left (  3\beta\left(\frac{m_{P}}{M}\right)y^2 z^2  - 5x_{\min }z^4 \right)^2  + \left (2 \beta \left(\frac{m_P}{M}\right) yz^3 \right)^2 \nonumber\\
& - & \kappa_S \left(\frac{M}{m_P}\right)^{2} x_{\min }^2 + \kappa_{h} \left(\frac{M}{m_P}\right)^{2} z^2 +\gamma\left(\frac{M}{m_P}\right)^{2} y^2 +\delta \left(\frac{M}{m_P}\right)^{4} y^4+\cdots\Bigg), \qquad
\end{eqnarray}
where $x_{min}$ is given by Eq.~(\ref{eq:14}). This two-field potential is displayed in Fig.~\ref{Fig1} for values of the various parameters given in the caption. A smooth trajectory suitable for inflation is clearly visible in this figure. For greater clarity a closer look at this trajectory is displayed in Fig.~\ref{Fig2}. The smooth trajectory here actually ends at a waterfall point which is shown by a red dot in Fig.~\ref{Fig2}. This is the reason why inflation along this trajectory is termed as pseudosmooth inflation \cite{Antusch:2012bp}. In this model the sneutrino $N$ field actually plays the role of the inflaton whereas variation in the $z$ field remains negligible during inflation.

\begin{figure}[t]
	\centering \includegraphics[width=10cm]{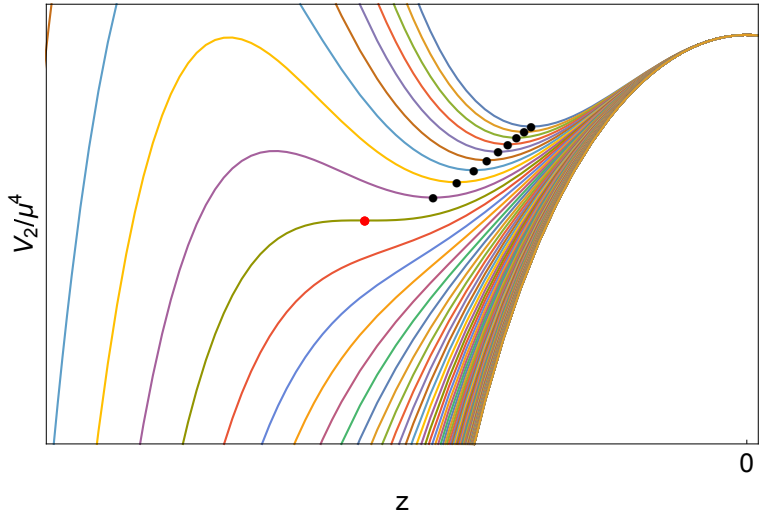}%
	\caption{The normalized two-field potential $V_{2}/\mu^4$ as a function of $z = \frac{h}{M}$ for various values of  $y = \frac{N}{M}$. We fix $\gamma = \delta = 0$, $\kappa_{S} = -\frac{1}{3}$, $ \kappa_{h} = -1$ and $M = 1\times 10 ^{17}$ GeV.}
	\label{Fig2}
\end{figure}
\subsection*{Waterfall Critical Point}
In pseudosmooth tribrid model inflation ends by a waterfall transition whereas in standard smooth hybrid model it ends by a slow-roll breaking with no waterfall along the complete smooth trajectory. The waterfall critical point $(z_{c},y_{c})$ can be obtained from the following conditions,
\begin{equation}\label{eq:19}
\begin{split}
\frac{\partial V_{2}(z_{c},y_{c})}{\partial z_{c}}=\frac{\partial^{2} V_{2}(z_{c},y_{c})}{\partial z_{c}^{2}}=0.
\end{split}
\end{equation}
Applying these conditions we obtain the following critical point, 
\begin{equation}\label{eq:21a}
y_{c}=\left(\frac{5}{2^\frac{5}{2}}\right)^\frac{1}{6}\left( \frac{(-\kappa_h)^\frac{1}{3} }{3\beta ^2
			 \left(\frac{
				M}{m_P}\right)^{4/3}}\right)^\frac{1}{4}\left(\frac{M}{m_P}\right),\qquad z_{c} = \left(\frac{2}{5}\right)^\frac{1}{3}\left((-\kappa_{h})\left(\frac{ M }{m_{P}}\right)^2\right)^\frac{1}{3},
\end{equation}
which also defines the condition for the end of inflation.
\subsection*{Effective Single-Field Potential}
Finally, minimizing $V_{2}$ with respect to field $z$ gives us the following approximate form of the  effective single-field potential,
\begin{equation}\label{eq:17}
V(y) \equiv V_{2}(y,z_{min}) \simeq \mu ^4 \left(1-\frac{1 }{18 
	y^4}\left(\frac{M}{m_{P}}\right)^{6}\left(\frac{\kappa_h}{\beta }\right)^{2}+\gamma \left(\frac{M}{m_{P}}\right)^{2}y^{2}+\delta\left(\frac{M}{m_{P}}\right)^{4}y^{4}\right),
\end{equation}
along the pseudosmooth trajectory approximately given by 
\begin{equation}\label{eq:18}
z_{min} \simeq \frac{1}{3}\sqrt{\frac{M^4 (-\kappa _{h })}{2\beta ^2 m_{P}^4
			y^4}}.
\end{equation}
In the leading order slow-roll approximation defined below, the prediction for the various inflationary parameters can be calculated by employing the above form of the potential. 

As natural values of the parameters are mostly involved in our analysis with field values of order GUT scale, a quick stabilization of all three phases is generally expected owing to their large masses. In the remaining 3-dimensional field space, any local minimum trajectory, if present, is expected to be isolated from the pseudosmooth trajectory and the global susy minimum. Any impact on inflationary predictions could be from the possible quantum tunneling transitions among these minima. A full analysis of estimating this impact on the model predictions can be quite cumbersome and lies beyond the scope of the current paper.

We have explicitly checked that the radiative corrections are negligibly small in our model. To see it with an order of magnitude estimate, we consider the following values of mass squared,
\begin{equation}
 4\left( (-\kappa_h) \pm \sqrt{ (-\kappa_h )^2 + \frac{2\kappa_h^3}{9 \beta^2 y^6} \left(\frac{M}{m_P}\right)^4} \right) \frac{\mu^4}{m_P^2}, \quad  2 (-\kappa_h)\frac{\mu^4}{m_P^2},\quad  2 (-\kappa_h)\frac{\mu^4}{m_P^2},
\end{equation}
for the inflaton-Higgs system in the limit $x = x_{\text{min}}\ll 1$ and $z=z_{\text{min}}\ll 1$. With an approppriate choice of renormalization scale, the radiative correction is proportional to $(-\kappa_h)^2(\mu/m_P)^4\mu^4$. As the quadratic mass term in the above potential plays equally important role in realizing inflation along with the other terms, the radiative  correction can be ignored compared to this term for $\gamma \gg (-\kappa_h)^2(\mu/m_P)^2(\mu/M)^2$ with $N = M$. This constraint is naturally satisfied in our numerical estimates. [Also see \cite{Antusch:2012jc} for a discussion of the smallness of radiative corrections in a typical model of tribrid inflation.]

The suppression of soft SUSY breaking terms with TeV scale soft masses is a common feature of tribrid inflation. In our model this can be seen with the following argument. As both $x_{\text{min}}$ and $ z_{\text{min}}$ are very small during the bulk of the inflationary phase ($10^{-10}\lesssim x_{\text{min}}\lesssim 10^{-5}$ and $0.005 \lesssim z_{\text{min}} \lesssim 0.05$), both $W$ and $z_i\partial W/ \partial z_i \sim \mathcal{O}(W)$ turn out to be negligibly small. Hence, the soft SUSY breaking $A$-term is negligible. Furthermore, the soft mass term $m_{\text{soft}}^2 |N|^2$ can be ignored compared to the quadratic mass term in the above potential for $m_{\text{soft}} \ll \sqrt{\gamma}\mu(\mu/m_P)$.  With $\mu \lesssim (10^{13}-10^{16})$ GeV, we obtain a soft mass $m_{\text{soft}} \ll \sqrt{\gamma} (10^5-10^{11})$ TeV. Thus, the approximation of ignoring the soft SUSY breaking terms throughout our calculations is justified.
\section*{Inflationary Slow-roll Parameters}
The slow-roll parameters are given below 
\begin{equation}\label{eq:25}
\epsilon(y) =\frac{1}{4}\left(\frac{m_{P}}{M}\right)^{2}\left(\frac{\partial_{y}V}{V}\right)^{2}, \, \eta(y)=\frac{1}{2}\left(\frac{m_{P}}{M}\right)^{2}\left(\frac{\partial_{y}^{2}V}{V}\right), \, \xi^{2}(y)= \frac{1}{4}\left(\frac{m_{P}}{M}\right)^{4}\left(\frac{\partial_{y}V \partial_{y}^{3}V}{V^{2}}\right), \notag
\end{equation}
where the subscript $y$ on $\partial$ denotes the derivative with respect to $y$. In the leading order slow-roll approximation, with $(\epsilon , \, \eta, \, \xi^{2}) \ll 1$, the tensor-to-scalar ratio $r$, the scalar spectral index $n_{s}$ and the running of the scalar spectral index $dn_{s}/d lnk$ are given by
\begin{eqnarray}
n_{s} &\simeq & 1 + 2\eta (y_0)- 6 \epsilon (y_0), \quad r \simeq  16\epsilon (y_0),  \\
\frac{dn_{s}}{d\ln k} &\simeq & 16 \epsilon (y_0) \eta (y_0)- 24 \epsilon^{2}(y_0) - 2 \xi^{2}(y_0),
\end{eqnarray}\label{eq:28}
where $y_0$ is the field value at the pivot scale which is taken to be at $k_{0} = 0.05\text{ Mpc}^{-1}$. The amplitude of curvature perturbation is given by
\begin{align}
A_{s}(k_{0})=\frac{1}{24\pi^{2}}\left. \left(\frac{V(y)/m_{P}^{4}}{\epsilon(y)}\right)\right|_{y=y_{0}},
\end{align}
where $ A_{s}(k_{0}) = 2.142 \times 10^{-9}$ is the Planck normalization at $k_{0} = 0.05\text{ Mpc}^{-1}$ \cite{Aghanim:2018eyx,Akrami:2018odb}. This constraint can be used to express $\mu$ in terms of $r$,
\begin{equation}
\mu  \simeq \left(\frac{3A_{s}(k_{0})\pi^{2}r}{2}\right)^\frac{1}{4}m_{P}. \label{mu}
\end{equation}
The number of efolds, $\Delta N $, from the pivot scale to the end of inflation is given by
\begin{equation}\label{efold1}
\Delta N  = 2\left(\frac{M}{m_{P}}\right)^{2} \int_{y_{e}}^{y_{0}}\frac{V}{\partial_{y}V}dy, \\
\end{equation}
where the field value at the end of inflation is $y_e = y_c$. Assuming standard thermal history we express the number of e-folds, $\Delta N$, in terms of the reheat temperature, $T_r$, as
\begin{align}\label{efold2}
\Delta N \simeq 47 + \frac{1}{3} \ln \left(\frac{T_r}{10^{6}\text{ GeV}}\right)+\frac{2}{3} \ln
\left(\frac{\mu}{10^{13}\text{ GeV}}\right).
\end{align}
In estimating the numerical predictions of the various inflationary parameters we set $T_r = 10^6$ GeV.
The realization of such a low reheat temperature and related non-thermal leptogenesis is justified after the discussion of numerical results. 
 %%%%%%%%%%%%%%%%%%%%%%%%%%%%%%%%%%%%
 %%%%%%%%%%%%%%%%%%%%%%%%%%%%%%%%%%%%
\section{Discussion of Numerical Results}
The numerical predictions of inflationary parameters are estimated by fixing the scalar spectral index at its central value, $n_s = 0.968$, and by setting $y_e = y_c = 1$. Using $y_c = 1$ in Eq.~(\ref{eq:21a}) the parameter $\beta$ can be written in terms of the other parameters as,
\begin{eqnarray} \label{eq:31}
\beta=\left(\frac{5}{2^\frac{5}{2}}\right)^\frac{1}{3}\left( \frac{(-\kappa_h)^\frac{1}{3} }{3	\left(\frac{
		M}{m_P}\right)^{4/3}}\right)^\frac{1}{2}\left(\frac{M}{m_P}\right)^2.
\end{eqnarray}
We fix the reheat temperature to its lowest possible value, i.e. $T_r = 10^{6}$ GeV, allowed by the successful non-thermal leptogenesis, as discussed in section 5 below. This value avoids the gravitino problem for a relatively wider range of gravitino mass \cite{Kawasaki:2004qu,Kawasaki:2008qe,Kawasaki:2017bqm}. We further impose $N_{0} \leq m_P$ as required by the reliability of supergravity corrections.

\begin{figure}[t]
	\includegraphics[width=8cm]{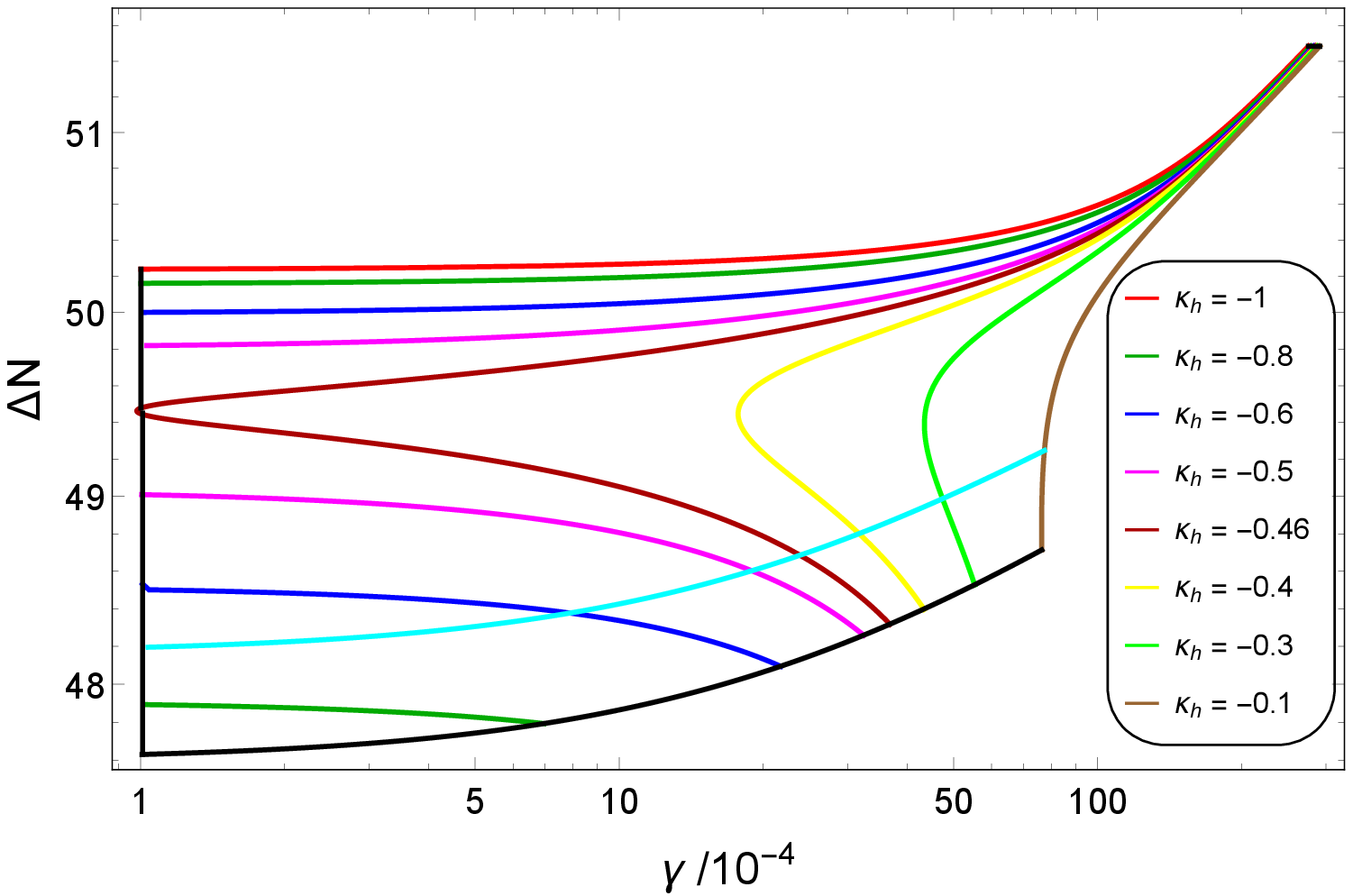}
	\includegraphics[width=8cm]{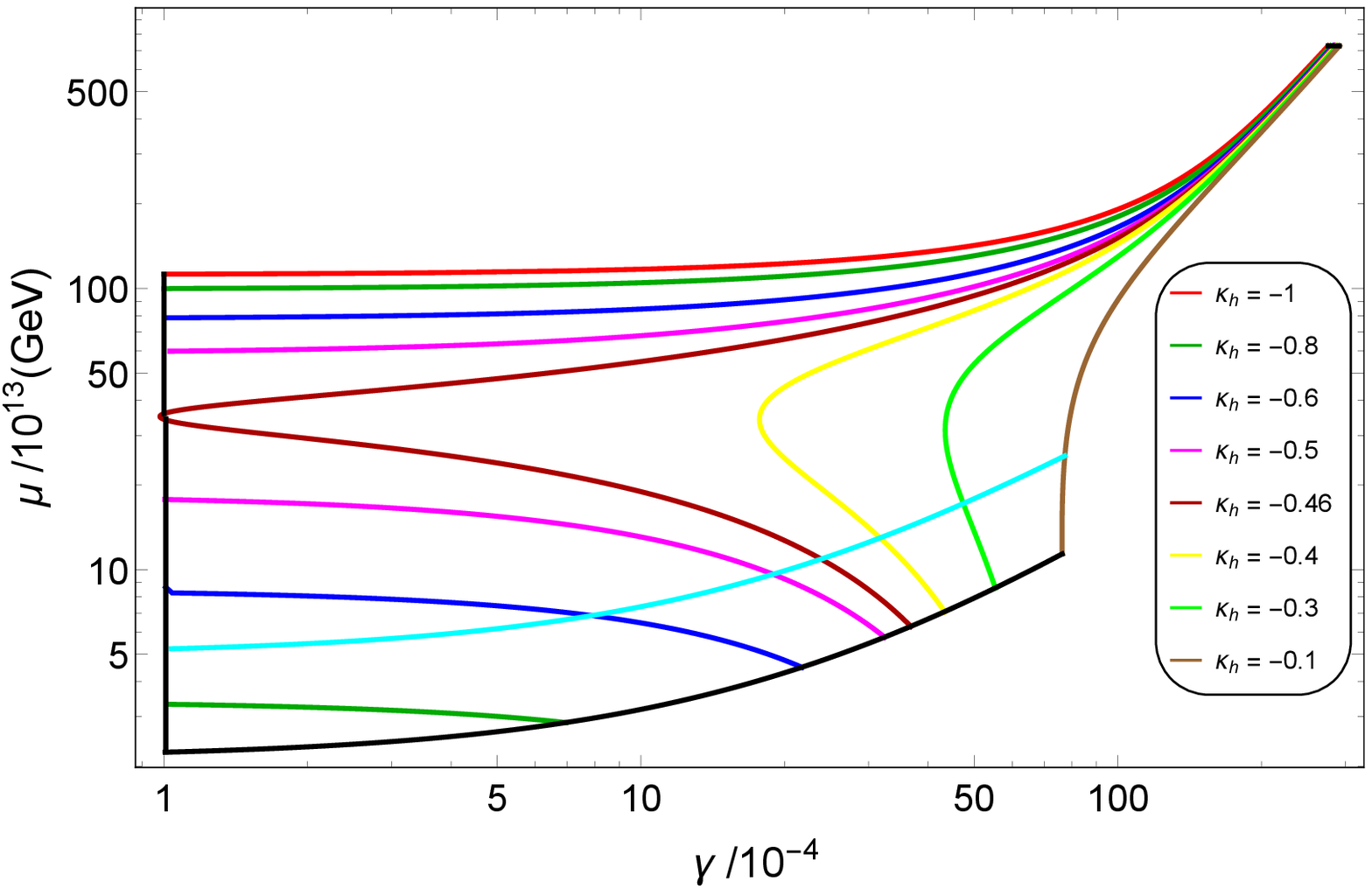}
	\caption{The number of e-folds $\Delta N$  (left panel) and the $\mu$ (right panel) versus the coupling  $\gamma$. We set the scalar spectral index,  $n_{s} = 0.968$ (central value of Planck's data), the end of inflation, $y_{e} =\frac{N_{e}}{M} = 1$, and the reheat temperature, $ T_{r} = 10^{6}$ GeV.}
	\label{Fig3}
\end{figure}

\begin{figure}[t]	
	\includegraphics[width=8cm]{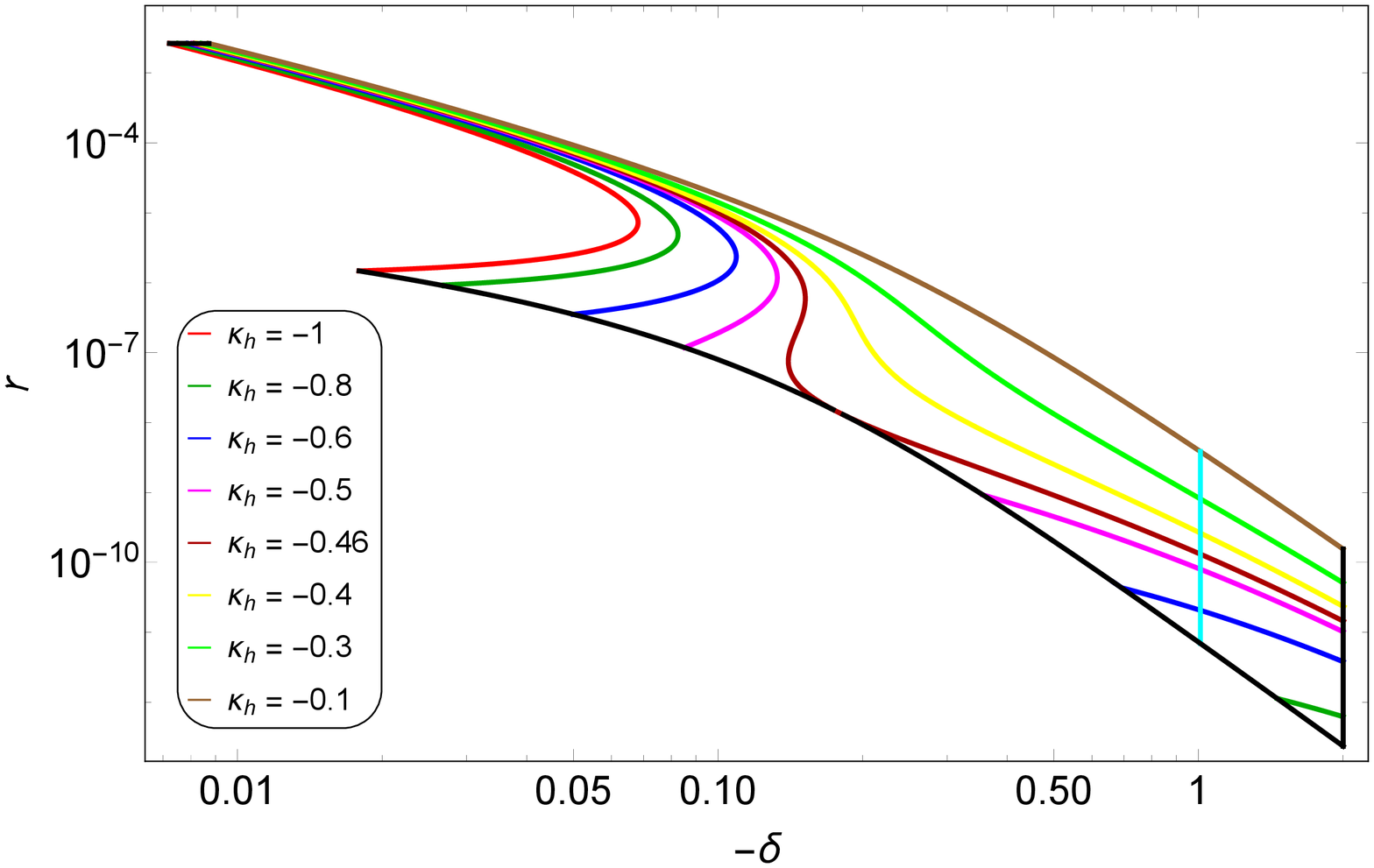}\,
	\includegraphics[width=8cm]{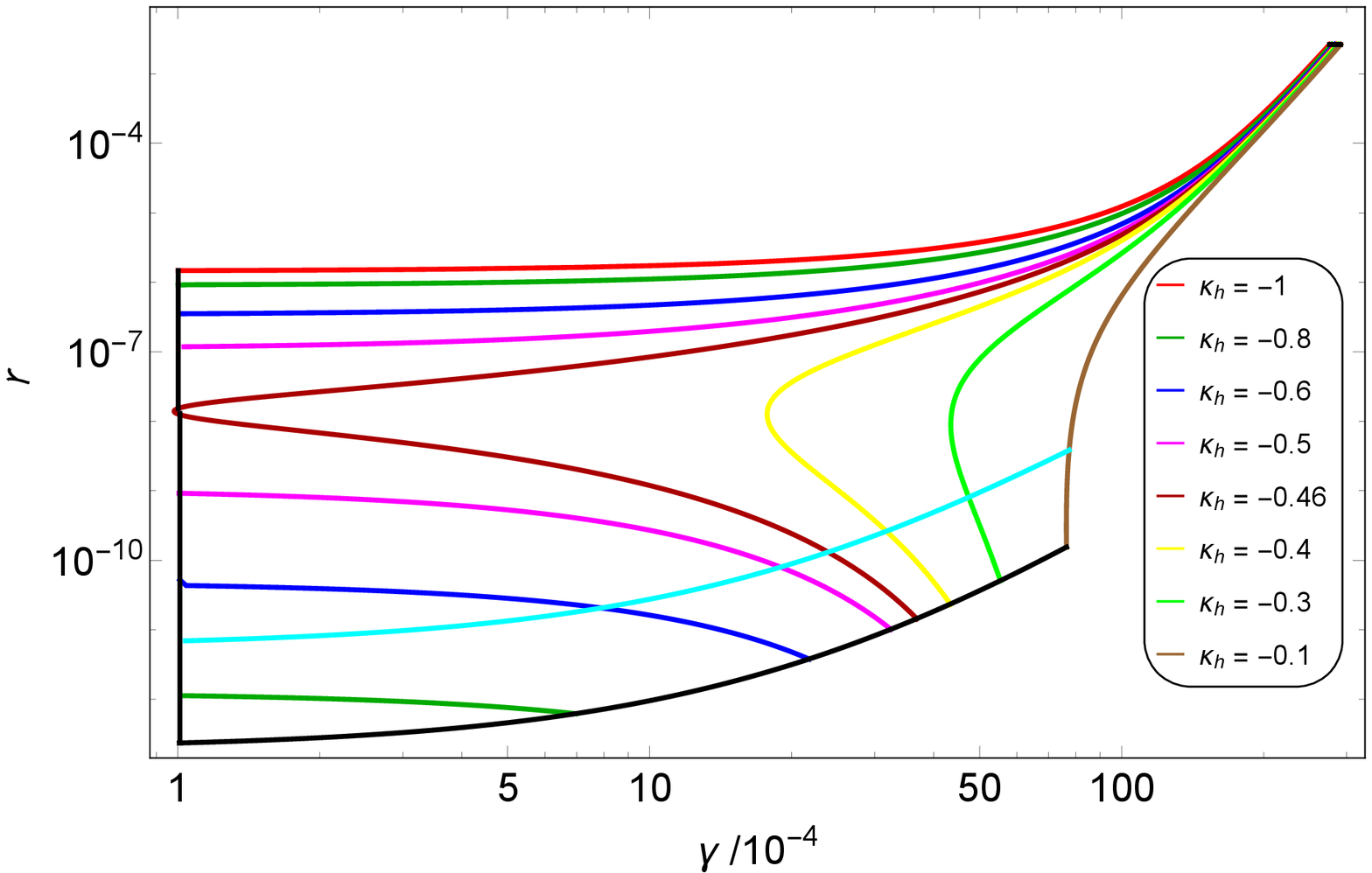}
	\caption{The tensor to scalar ratio $r$  versus the coupling $\gamma$ (left panel) and the coupling $\delta$ (right panel). We set the scalar spectral index  $n_{s} = 0.968$ (central value of Planck's data), the end of inflation $y_{e} =\frac{N_{e}}{M} = 1$ and the reheat temperature $ T_r = 10^{6}$ GeV.}
	\label{Fig4}
\end{figure}

To identify a relatively natural region of parametric space we restrict $ |\gamma| \gtrsim 10^{-4}$, $|\delta| \lesssim 1$ and $-1\leq \kappa_h \leq -0.1$. We require $\kappa_h<0$ in order to generate a smooth inflationary track while a successful realization of inflation further requires $\delta<0$ and $\gamma>0$. Assuming $\alpha \ll 13/30$, the parameter $\mu$ and tensor to scalar ratio, $r$, can be directly related to $M$, via Eqs.~(\ref{min}) and (\ref{mu}), as
\begin{equation}
\mu \simeq \sqrt{\frac{13}{240\sqrt{15}}} \left(\frac{M}{m_P}\right)^{5/2} m_P,\qquad  r \simeq \left(\frac{1}{2\times 10^{-4}}\right)\left(\frac{M}{m_{P}}\right)^{10}.
\end{equation}
From Eq.~(\ref{efold2}), we can also write down $\mu$ in terms of $\Delta N$ as $\mu/10^{13}\text{GeV} = e^{3(\Delta N - 47)/2}$ with $T_r =10^6$ GeV. After solving Eq.~(\ref{efold1}) and Eq.~(\ref{efold2}) numerically with the above mentioned constraints we obtain $\Delta N \simeq 47.6-51.5$ as shown in Fig.~\ref{Fig3}. This range of $\Delta N$ corresponds to the range $2 \lesssim \mu/10^{13} \text{GeV} \lesssim 744$. This yields, via the above equations, $ 5.4 \times 10^{16} \lesssim M/\text{GeV} \lesssim  5.6 \times 10^{17}$ and $ 10^{-13}  \lesssim r  \lesssim 2.4\times 10^{-3}$. Moreover, the approximate upper bound $M \lesssim 5.6 \times 10^{17}$ with $N_0 = m_P$ translates into the upper bound $y_0 \lesssim 4 $. These approximate estimates are compatible with the exact numerical results displayed in Fig.~\ref{Fig4}. 


\begin{figure}[t]
\includegraphics[width=8cm]{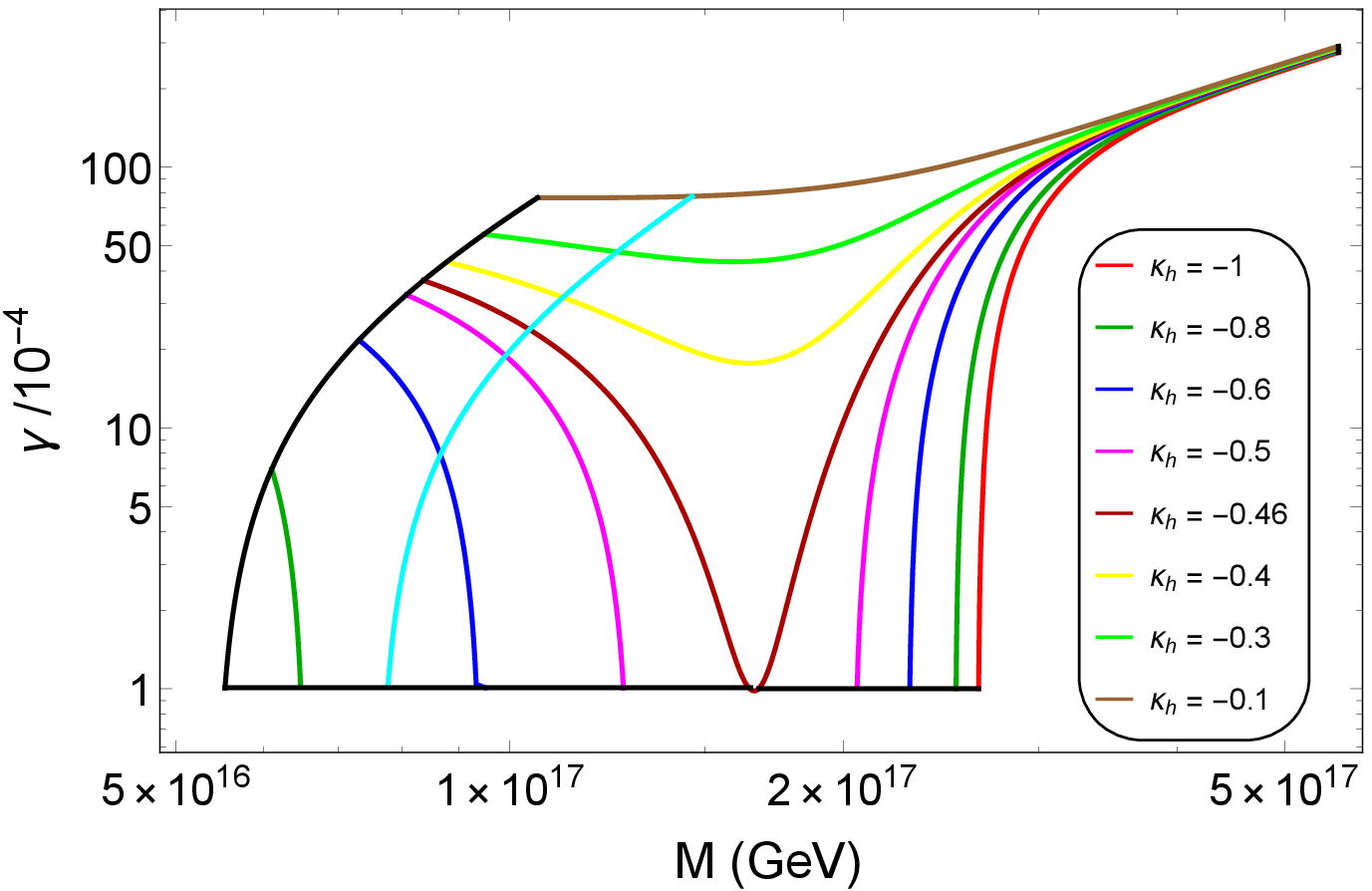}
\includegraphics[width=8cm]{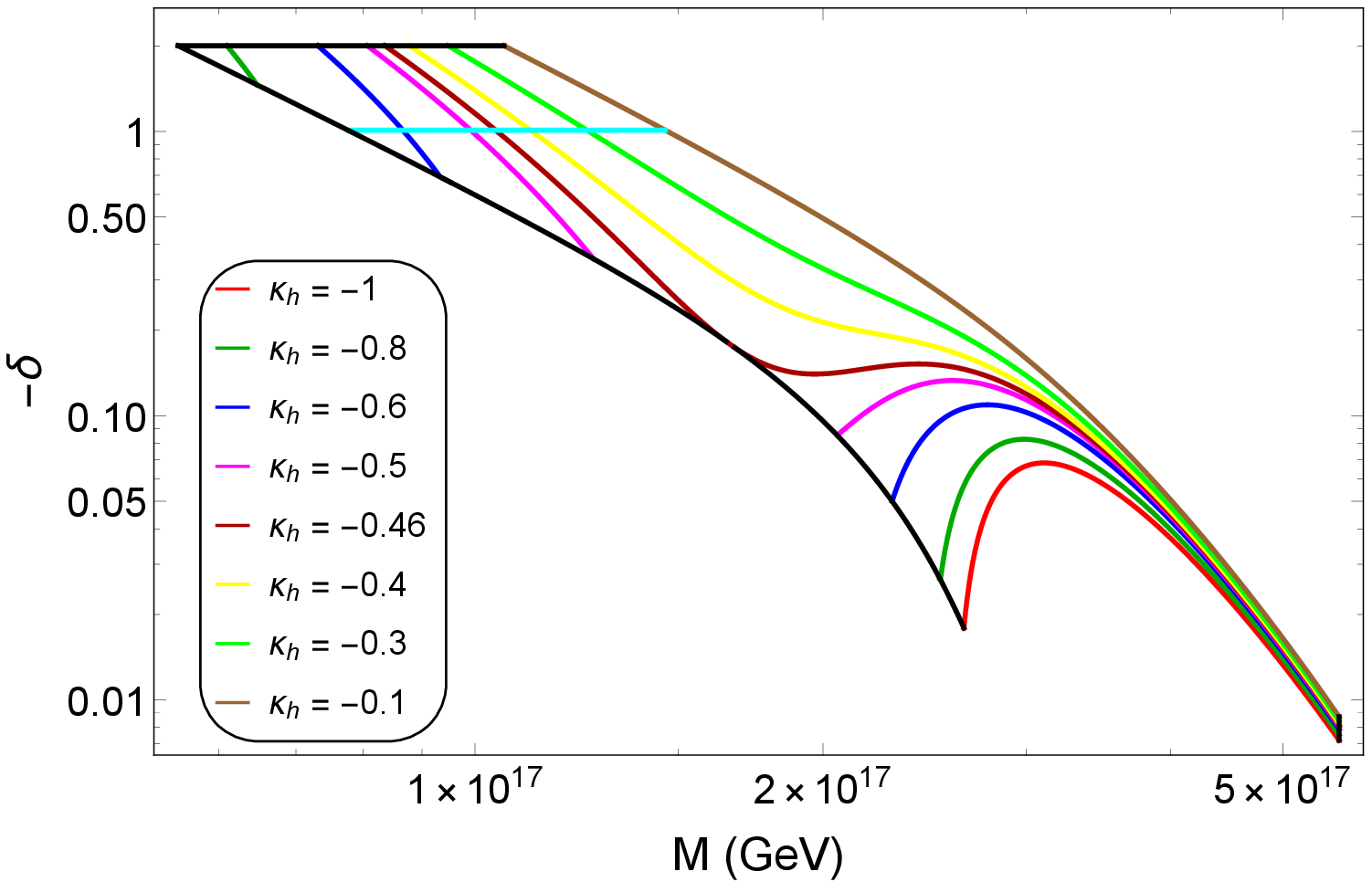}
	\caption{. The coupling  $\gamma$ (left panel) and the coupling $\delta$ (right panel) versus the gauge symmetry breaking scale $M$. We set the scalar spectral index  $n_{s} = 0.968$ (central value of Planck's data), the end of inflation $y_{e} =\frac{N_{e}}{M} = 1$ and the reheat temperature $ T_{r}= 10^{6}$ GeV.}
	\label{Fig5}
\end{figure}

 The explicit dependence of $\gamma$ and $\delta$ in terms of the remaining parameters can be obtained from Eq.~(\ref{eq:25}) as
\begin{eqnarray}
\gamma&\simeq&\frac{1}{4}\left(1-n_{s}\right)+\frac{3\sqrt{r}}{8 y_{0} }\left(\frac{m_P}{M}\right)+\frac{8\kappa_h}{3y_{0}^6 }\left(-\frac{2\kappa_h}{5}\left(\frac{M}{m_P}\right)^2\right)^\frac{2}{3},\label{eq:33} \\
\delta&\simeq&\frac{(n_{s}-1)}{8 y_{0}^2}\left(\frac{m_P}{M}\right)^2 - \frac{\sqrt{r}}{16 y_{0}^3}\left(\frac{m_P}{M}\right)^3-\frac{\kappa_h}{y_{0}^8}\left(-\frac{2\kappa_h}{5}\left(\frac{m_P}{M}\right)\right)^\frac{2}{3}. \label{eq:34}
\end{eqnarray}
For relatively large values of $r$ only the first two terms in the above expressions are important. This leads to the weak dependence of $\gamma$ and $\delta$ on $\kappa_h$ as depicted in Figs.~\ref{Fig3}-\ref{Fig5}. For instance, with $N_0 = m_P$ we obtain
\begin{eqnarray}
\gamma&\simeq&\frac{1}{4}\left(1-n_{s}\right)+\frac{3\sqrt{r}}{8}\simeq 0.026, \quad \delta \simeq -\frac{(1-n_{s})}{8} - \frac{\sqrt{r}}{16} \simeq -0.007. 
\end{eqnarray}
This again is a very good approximation of the more precise numerical estimates shown in Figs.~\ref{Fig3}-\ref{Fig5}. 

\begin{figure}[H]
	\includegraphics[width=8cm]{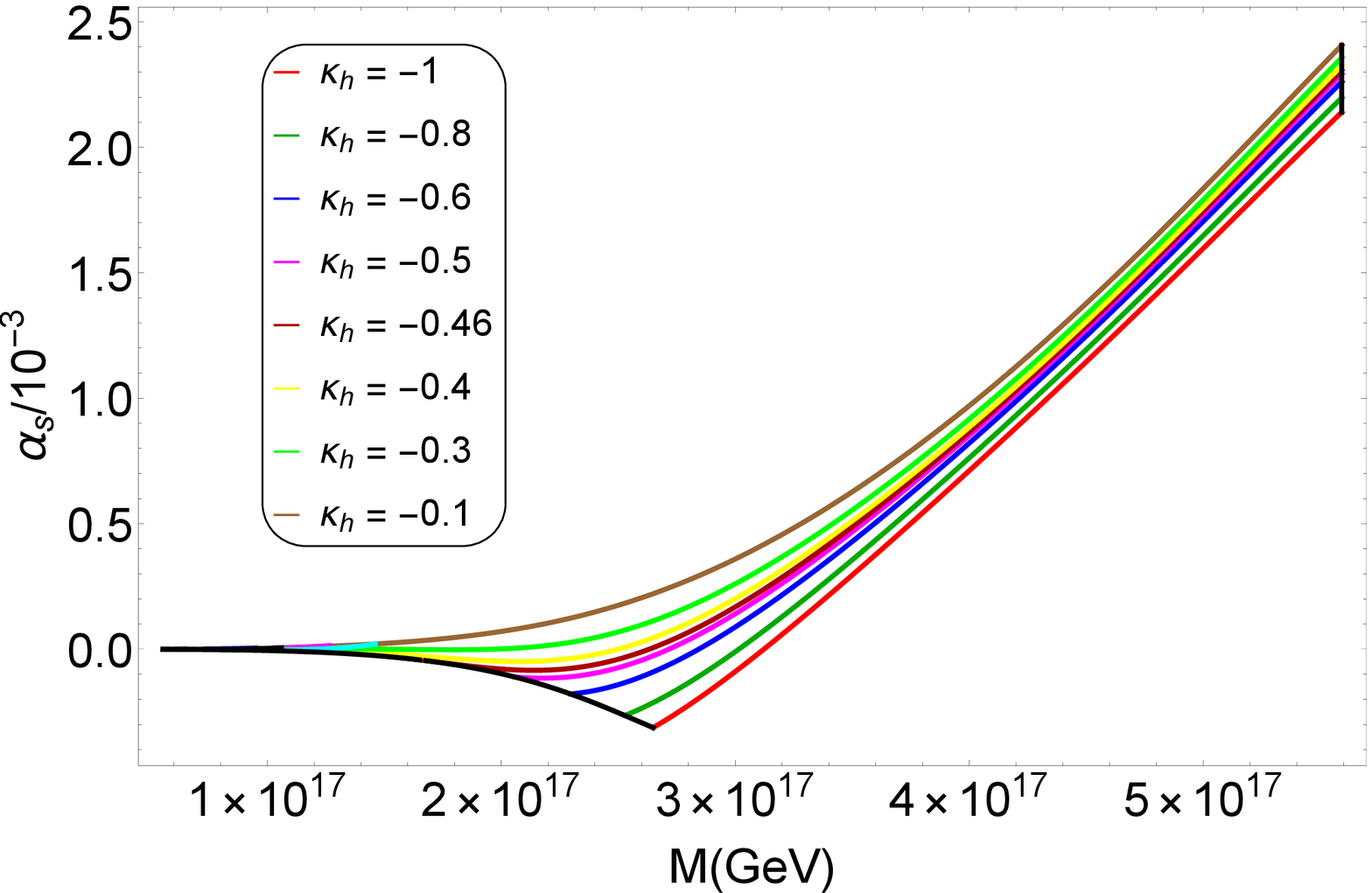}\,
	\includegraphics[width=7.7cm]{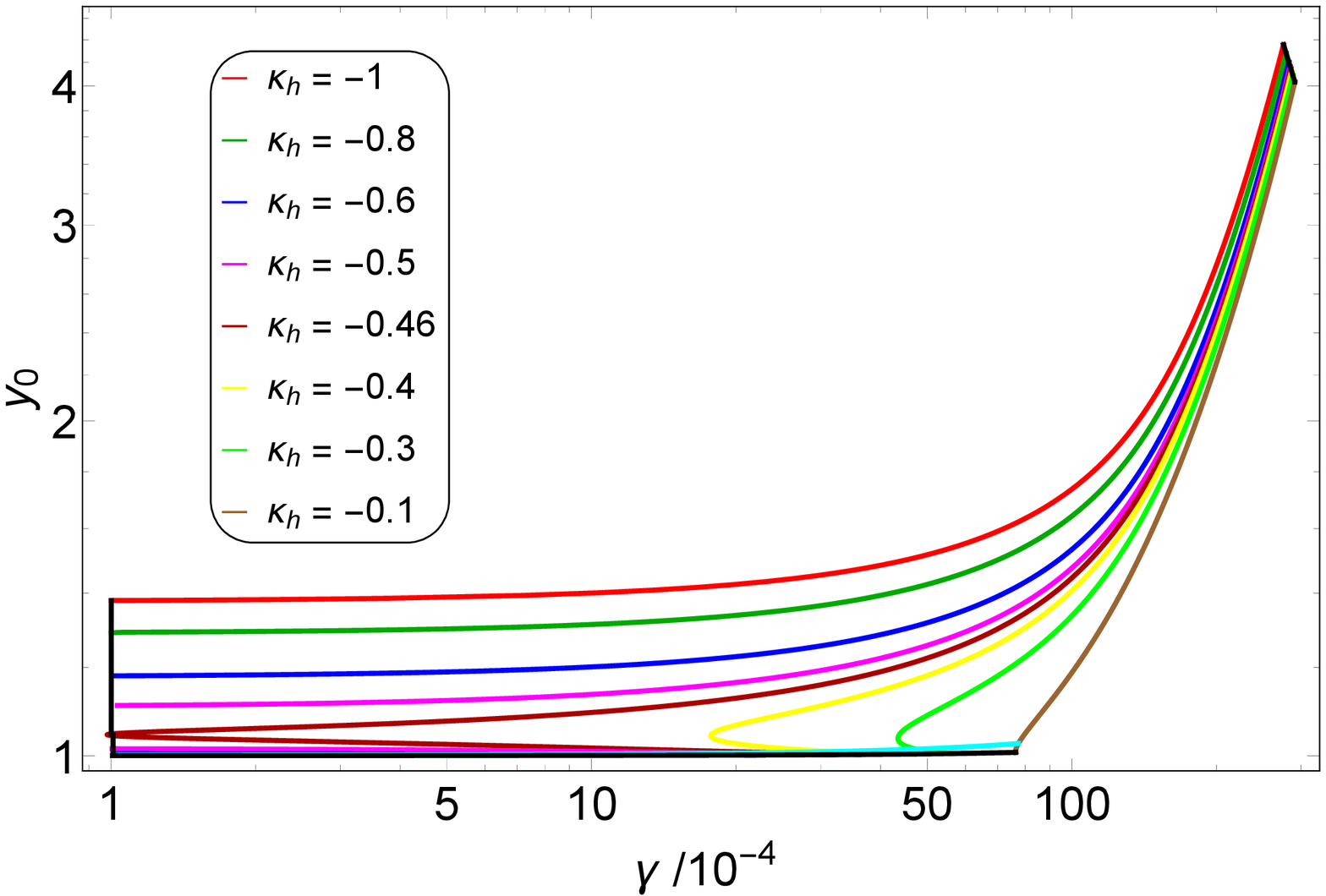}
	\caption{The running of spectral index  $\alpha_s \equiv \frac{dn_{s}}{d\ln k}$  (left panel) and the normalized field value $y_{0} = \frac{N_{0}}{M}$ (right panel)  versus the coupling $\gamma$.  We set the scalar spectral index  $n_{s} = 0.968$ (central value of Planck's data), the end of inflation $y_{e} =\frac{N_{e}}{M} = 1$ and the reheat temperature $ T_{r} = 10^{6}$ GeV.}
	\label{Fig6}
\end{figure}

In the small $r$ limit the second terms in Eqs.~(\ref{eq:33}) and (\ref{eq:34}) become negligible. The last term in Eq.~(\ref{eq:33})  becomes comparable to the first term while making $\gamma$ small compared to $(1-n_s)/4$. This fact allows us to write $y_0$ and $M$ in terms of $\delta$ and $\kappa_h$ as 
\begin{equation}
y_{0} \simeq \left(\frac{2^7(-\kappa_{h})^5}{3^3\times5^2(n_{s}-1)\delta^2}\right)^{\frac{1}{22}},\quad 
M \simeq \frac{1}{4\delta^\frac{3}{2}}\left(\frac{3^\frac{3}{2}\times 5((n_{s}-1)\delta^{2})^{6}}{2^9(-\kappa_{h})^\frac{5}{2}}\right)^\frac{1}{11}m_P,
\end{equation}
for $\gamma \ll (1-n_s)/4$. Using these expressions with $y_0 \simeq 1$ and $-0.8 \lesssim \kappa_h \lesssim -0.5$ we obtain $-1.4 \lesssim \delta \lesssim -0.4$ and  $ 6.5 \times 10^{16} \text{ GeV } \lesssim M \lesssim 1.2 \times 10^{17} \text{ GeV} $, which is in good agreement with our numerical estimates, as shown in Figs.~\ref{Fig4}-\ref{Fig6}. Note that in these figures the lower bounds on $M$ and $r$ are very sensitive to the upper bound on $|\delta|$. For example, increasing the value of $|\delta|$ above unity, represented by the cyan curve, can further reduce the lower bound on $M$ and $r$ as shown by the $\delta = -2$ black curve. Moreover, the bound on $M$ with $ -1 \leq \kappa_h \leq -0.1$ gives the range, $10^{-3} \lesssim \beta \lesssim 10^{-2}$, via Eq.~(\ref{eq:31}).

Finally, the running of spectral index $\frac{dn_{s}}{d\ln k}$ can be described in terms of $r$, $n_s$ and other parameters as
\begin{equation}
\alpha_s \equiv \frac{dn_{s}}{d\ln k} \simeq  r \frac{(n_s -1)}{2} - \frac{3}{32} r^2 - 2 \sqrt{r} \left( \frac{2^{2/3}5^{1/3}(-\kappa_h)^{5/3}}{y_0^7}\left( \frac{M}{m_P}\right)^{1/3} + 3 y_0 \delta \left( \frac{M}{m_P}\right) \right).
\end{equation}
The largest possible value of $\frac{dn_{s}}{d\ln k} \lesssim 2 \times 10^{-3}$ appears in the large $r$ limit (see Fig.~\ref{Fig6}). This shows that our results are perfectly consistent with the latest Planck data results.
\section{Reheating and Non-Thermal Leptogenesis}
 The reheating in the current model proceeds in analogy with the $Z_4$ sneutrino model \cite{Antusch:2004hd}, and the observed baryon asymmetry is explained by nonthermal leptogenesis \cite{Lazarides:1991wu}. The inflaton $N_{1}$, being the lightest sneutrino field, is assumed to play a dominant role in reheating after inflation and subsequent leptogenesis. This is possible if the Higgs field decays earlier than the inflaton as discussed below. From Eq.~(\ref{eq:6}), the mass of the singlet sneutrino inflaton $N_{1}$ is given by
\begin{align}\label{eq:21}
&M_R^{I} = \left(\frac{\beta_{11}}{2\sqrt{15}} \right)\frac{ M^{3}}{m_{P}^{2}} =\left(\frac{13}{8 y_c^2}\right)\left( \frac{(-\kappa_h)^\frac{1}{3} }{3^4\times 2^\frac{5}{3} \times 5^\frac{7}{3} }\right)^\frac{1}{2}\left( \frac{M}{m_{P}}\right)^{10/3}M.
\end{align}
The inflaton decays through the effective Yukawa coupling,
\begin{eqnarray}
\lambda_{1j}^{\nu}\frac{Tr(24_{h}^{2})}{m_{P}^{2}}N_{1}\overline 5_{j}5_{h}+\frac{\tilde\lambda_{1j}^{\nu}}{m_{P}^{2}}N_{1}\overline 5_{j}24_{h}^{2}5_{h}\supset Y_{1j}^{\nu} N_{1}L_{j} H_{u},
\end{eqnarray}
into sleptons and Higgs or into lepton and Higgsino with a decay width given by  
\begin{align}\label{eq:22}
&\Gamma_{N_{1}} \simeq \frac{y_{\nu}^2}{4\pi} M_{R}^I = \frac{y_{\nu}^2}{4\pi} \left(\frac{13}{8 y_c^2}\right)\left( \frac{(-\kappa_h)^\frac{1}{3} }{3^4\times 2^\frac{5}{3} \times 5^\frac{7}{3} }\right)^\frac{1}{2} \left( \frac{M}{m_{P}}\right)^{10/3}M,
\end{align}
where,
\begin{align}\label{eq:23}
y_{\nu}^2 \equiv  (Y_{\nu}Y_{\nu}^{\dagger})_{11}, \quad Y_{ij}^{\nu}& =\left(\frac{\lambda_{ij}^{\nu}}{2} + \frac{\tilde\lambda_{ij}^{\nu}}{15}\right) \left(\frac{M}{m_{P}}\right)^{2}.
\end{align}

\begin{figure}[t]
		\includegraphics[width=8cm]{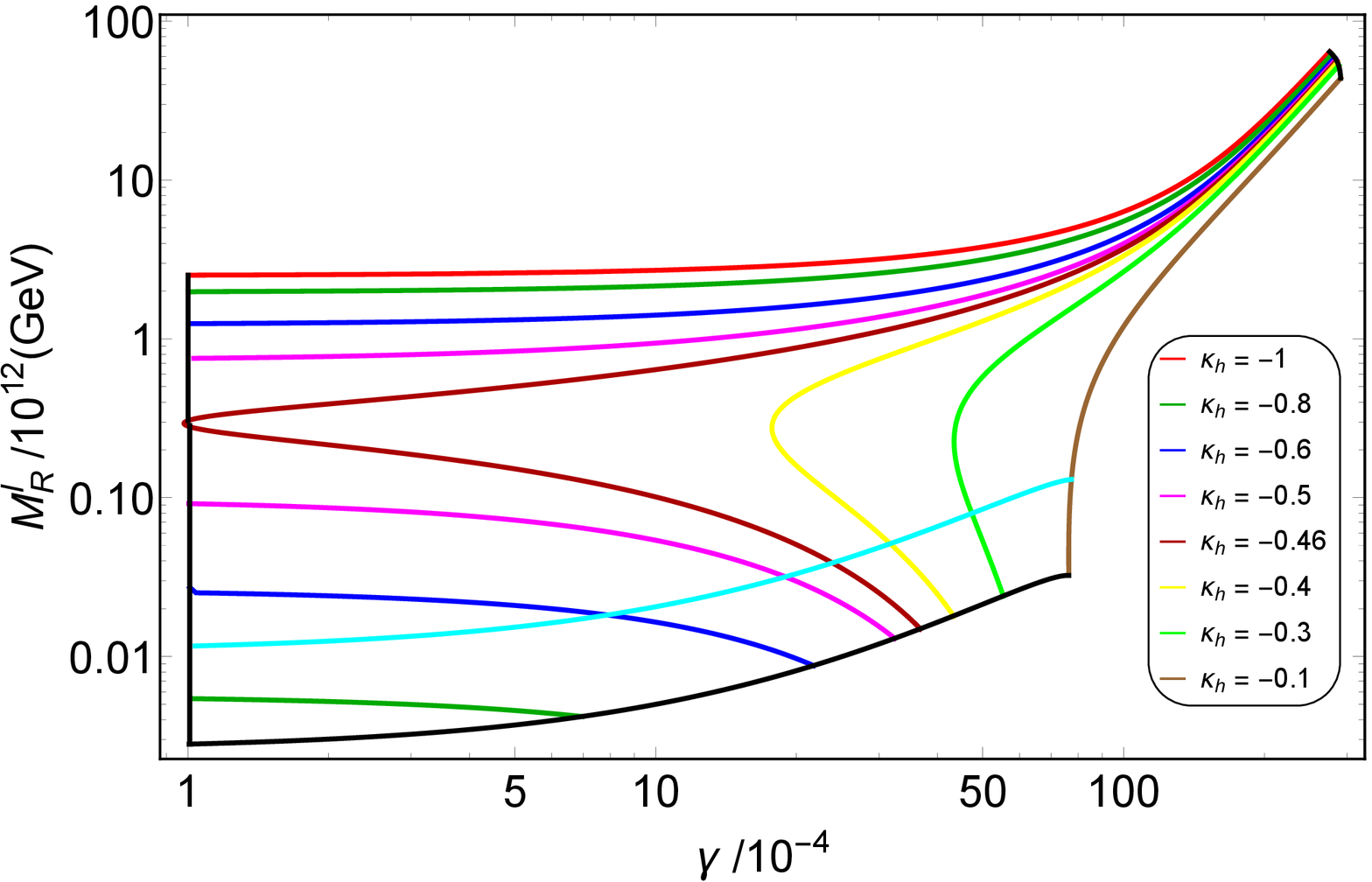}\,
	\includegraphics[width=8cm]{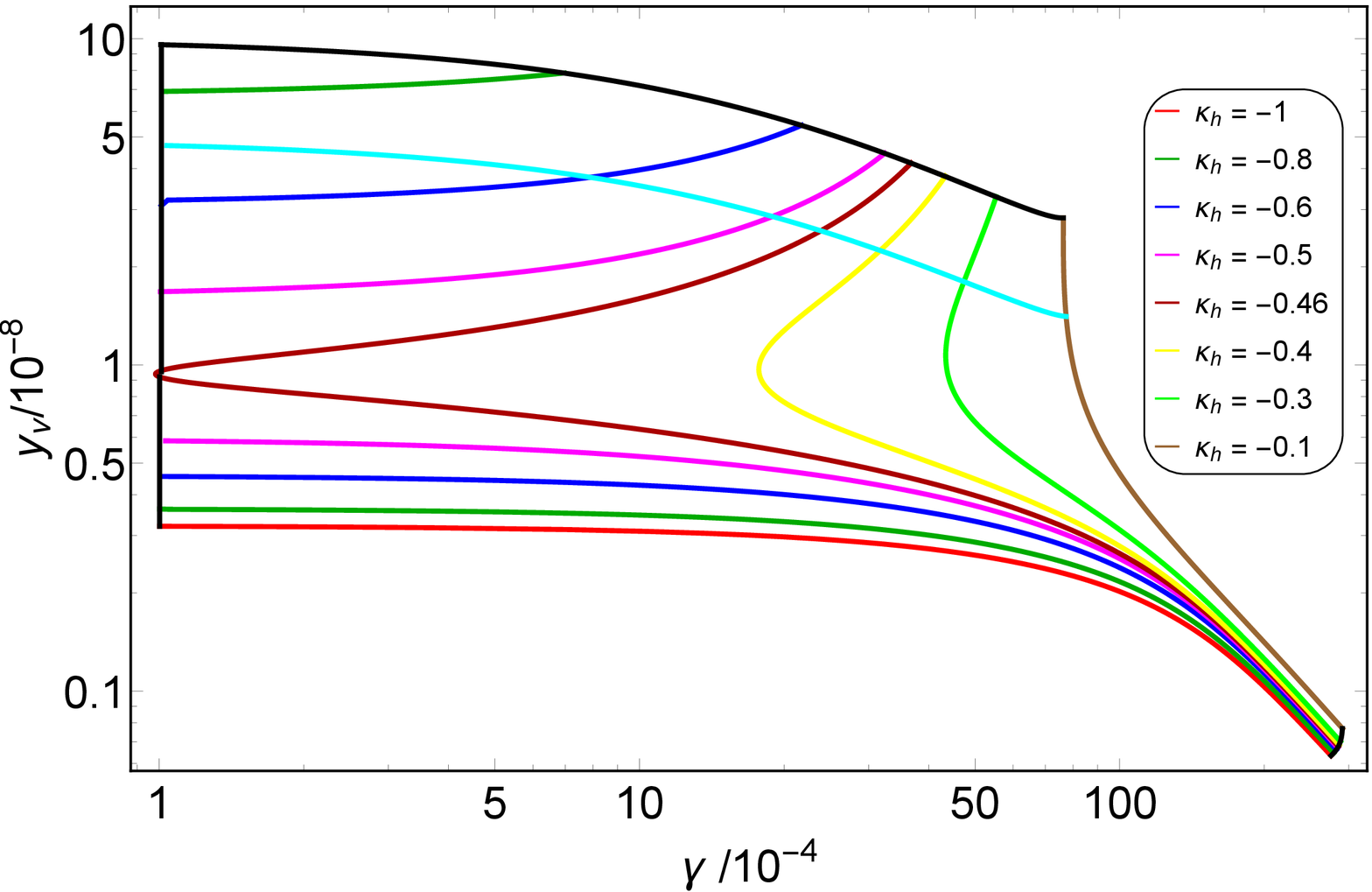}%
	\caption{ The mass of the (s)neutrino inflaton $M_{R}^I$ (left panel) and  the neutrino Yukawa coupling $y_{\nu}$ (right panel) versus the coupling $\gamma$. We set the scalar spectral index  $n_{s} = 0.968$ (central value of Planck's data), the end of inflation $y_{e} =\frac{N_{e}}{M} = 1$ and the reheat temperature $ T_{r} = 10^{6}$ GeV.}
	\label{Fig:7}
\end{figure}

Compared to the $Z_4$ sneutrino tribrid model, we have an extra suppression factor $(M/m_P)^2$ which can make the fundamental Yukawa couplings ($\lambda_{ij}^{\nu},\tilde\lambda_{ij}^{\nu}$) relatively natural. Assuming the Higgs decay rate to be larger than the inflaton decay rate we obtain the following bound on $y_{\nu}$,
\begin{equation}
y_{\nu}^2 \ll \left(\frac{M_R^{(2,3)}}{M} \right) \left(\frac{M_R^{(2,3)}}{M_R^I}\right),
\end{equation}
where $M_R^{(2,3)}$ are the masses of the heavier neutrinos $N_{(2,3)}$. This bound is easily satisfied in our model for the numerical data displayed in Fig.~\ref{Fig:7}.
After inflation, the universe reheats via inflaton decay to a temperature,
\begin{equation}
T_{r} \simeq \left( \frac{90}{g_* \pi^{2}} \right)^\frac{1}{4}\sqrt{\Gamma_{N_{1}}m_{P}},
\end{equation}
where $g_* =228.75$. 

The lepton asymmetry generated by the inflaton decay can be partially converted into the observed baryon asymmetry through sphaleron processes. We assume $M_{R}^I\gg T_{r}$ in order to suppress the washout factor of lepton asymmetry. The baryon asymmetry can be estimated in terms of the lepton asymmetry factor $\varepsilon_L$ as
\begin{equation}
\frac{n_{B}}{n_{\gamma}}\simeq -1.84 \,\varepsilon_L  \frac{T_{r}}{M_{R}^I},
\end{equation} 
where $\varepsilon_L$ satisfies the following bound,
\begin{equation}
(-\varepsilon_L)  \lesssim \frac{3}{8\pi}  \frac{\sqrt{\Delta m_{31}^{2}} M_{R}^I}{\langle H_{u}\rangle^{2}},
\end{equation} 
assuming a hierarchical structure of neutrino masses. Here, the atmospheric neutrino mass squared difference is $\Delta m_{31}^{2}\approx 2.6 \times 10^{-3}$ eV$^{2} $  and $\langle H_{u}\rangle = 174$ GeV in the large $\tan\beta$ limit. Finally, the bound on $\varepsilon_L$ translates into the bound on reheat temperature $T_r \gtrsim 10^6$ GeV for the observed baryon-to-photon ratio  $n_\mathrm{B} /n_\gamma = (6.10 \pm 0.04) \times 10^{-10}$ \cite{Cyburt:2015mya}. Thus, the reheat temperature is small enough to avoid the gravitino problem. We set $T_r = 10^6$ GeV in all numerical work and obtain $ 2.8 \times 10^{9} \lesssim M_R^I / \text{GeV} \lesssim 4 \times 10^{13}$ and $ 6.3 \times 10^{-10} \lesssim y_{\nu} \lesssim 10^{-7}$, as depicted in the Fig.(\ref{Fig:7}).

\section{Gauge Coupling Unification in R-symmetric SU(5)}
According to the no-go theorem mentioned in \cite{Barr:2005xya,Fallbacher:2011xg}, we obtain `massless' fields in any $R$-symmetric grand unified theory (GUT) based on a simple gauge group  after spontaneous breaking of the GUT symmetry. These fields, however, can acquire TeV scale masses from the soft SUSY breaking terms. In our $R$-symmetric $SU(5)$ model we obtain light ($\sim$ TeV) octet and triplet components from the $24_H$ Higgs field \cite{Khalil:2010cp}. The presence of these light fields, in turn, ruins the successful gauge coupling unification feature of MSSM. To circumvent this problem we add  copies of vectorlike families $5+\bar{5}+10+\bar{10}$. This does not solve the problem of gauge coupling unification unless we allow mass splitting within their MSSM field components, 
\begin{eqnarray}
5 + \overline{5} &=& \left( D + \overline{D}, L + \overline{L}\right), \quad 10 + \overline{10} = \left( Q + \overline{Q}, U + \overline{U}, E + \overline{E}\right).
\end{eqnarray}
This splitting is achieved in a way similar to the doublet-triplet splitting but with far less fine tuning. With additional vectorlike families we obtain the following mass terms in the superpotential,
\begin{eqnarray}
W &\supset& \frac{\lambda_{ij}^{(10,\overline{10})}}{m_{P}} Tr(24_{h}^{2}) Tr(10_i\overline{10}_j) +\frac{\tilde{\lambda}_{ij}^{(10,\overline{10})}}{m_{P}} Tr(10_i 24_{h}^{2}  \overline{10}_j) \\
 &+& \frac{\lambda_{ij}^{(5,\overline{5})}}{m_{P}} Tr(24_{h}^{2}) Tr(5_i\overline{5}_j) +\frac{\tilde{\lambda}_{ij}^{(5,\overline{5})}}{m_{P}} Tr(5_i 24_{h}^{2}  \overline{5}_j), \\
 &\supset& M_{Q} Q \overline{Q} + M_{U} U \overline{U} + M_{E} E \overline{E}
 + M_{D} D \overline{D} + M_{L} L \overline{L},
\end{eqnarray}
with $Z_5$-charge, $q_5 \left(5\,\overline{5},10\,\overline{10}\right)=(3,3)$, and $R$-charge, $R \left(5\,\overline{5},10\,\overline{10}\right) = (1,1)$. For simplicity, assuming $\lambda_{ij} = \delta_{ij}\lambda$ and $\tilde{\lambda}_{ij}= \delta_{ij}\tilde{\lambda}$, the masses of the MSSM field components of a vectorlike family are given by 
\begin{align}
M_{E}& = \frac{30\lambda^{(10,\overline{10})} + 9\tilde{\lambda}^{(10,\overline{10})}}{30}\left( \frac{M^{2}}{m_{P}}\right),\\
 M_{Q }& = \frac{60\lambda^{(10,\overline{10})} + 13\tilde{\lambda}^{(10,\overline{10})}}{30}\left( \frac{M^{2}}{m_{P}}\right),\\
 M_{U }&= \frac{15\lambda^{(10,\overline{10})} + 2\tilde{\lambda}^{(10,\overline{10})}}{30}\left( \frac{M^{2}}{m_{P}}\right),\\
 M_{D }& = \frac{15\lambda^{(5,\overline{5})}+2\tilde{\lambda}^{(5,\overline{5})}}{30}\left( \frac{M^{2}}{m_{P}}\right),\\
 M_{L }& = \frac{20\lambda^{(5,\overline{5})} + 6\tilde{\lambda}^{(5,\overline{5})}}{40}\left( \frac{M^{2}}{m_{P}}\right).
\end{align}
\begin{figure}[t]
	\centering
     \includegraphics[width=8cm]{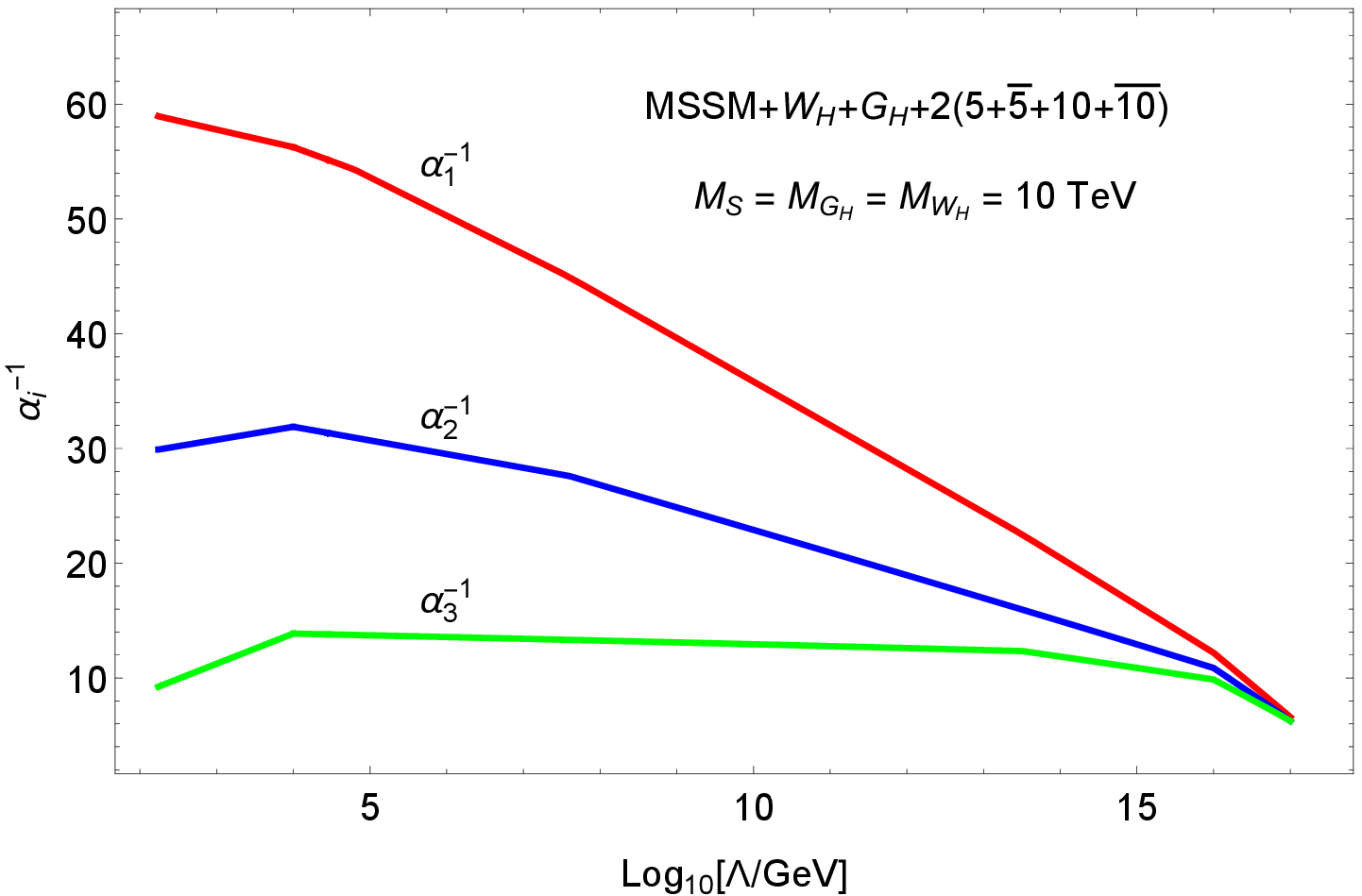} %
    \includegraphics[width=8cm]{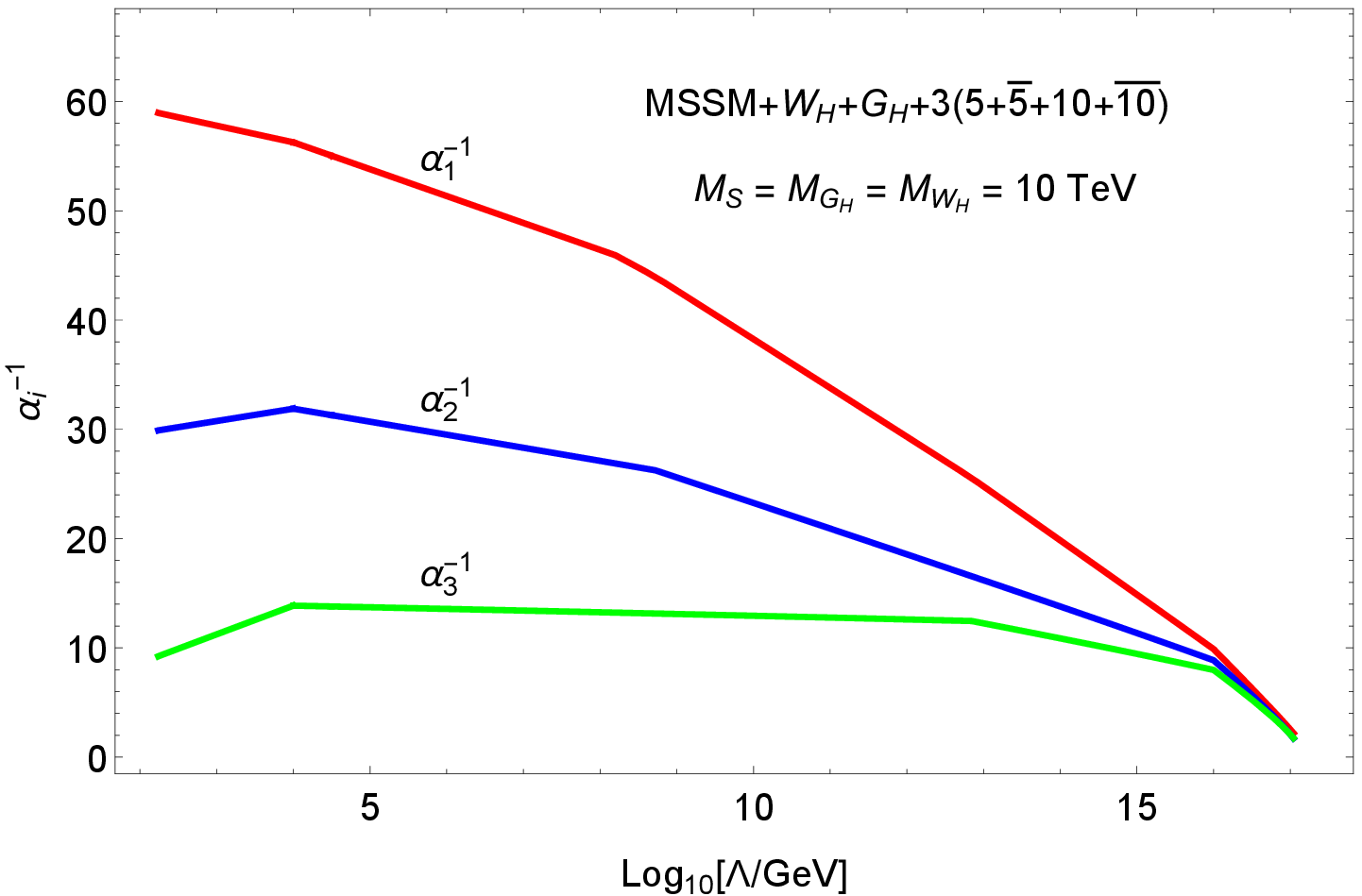}%
	\caption{The evolution of the inverse gauge couplings versus the energy scale $\Lambda$ in $R$-symmetric $SU(5)$ model, with two (left-panel) and three (right-panel) generations of vectorlike families $(5+ \overline{5} + 10 + \overline{10} )$. The effective SUSY breaking scale is set at $M_S = 10 $ TeV. The masses of  vectorlike MSSM components are taken as $M_{Q} = M_{U} = 10^{16}$ GeV, $M_{D} = 10^{13.5}$ GeV, $M_{L} = 10^{7.6}$ GeV, $M_{E} = 10^{4.8}$ GeV (left-panel) and  $M_{Q} = M_{U} = 10^{16}$ GeV, $M_{D} = 10^{12.845}$ GeV, $M_{L} = 10^{8.719}$ GeV, $M_{E} = 10^{8.2}$ GeV (right-panel). The GUT scale, $M_{\text{GUT}}=10^{17}$ GeV, in both cases.}
		\label{fig:phivsns}
\end{figure} 
Now we can make a selected single field component to be light in both vectorlike multiplets of $SU(5)$. We choose light masses for $E+\overline{E},$ and $L+\overline{L}$ with $\left( 30\lambda^{(10,\overline{10})} + 9\tilde{\lambda}^{(10,\overline{10})}\right) \sim 0$, and $\left( 20\lambda^{(5,\overline{5})} + 6\tilde{\lambda}^{(5,\overline{5})} \right) \sim 0$. The other components can have masses as large as $M^{2}/m_{P} \sim 10^{16}$ GeV.  A successful gauge coupling unification can be achieved with two or three generations of additional vectorlike families. This is shown in Fig.~\ref{fig:phivsns} with different mass splitting patterns described in its caption. As the triplet and the octet components of $24_H$ Higgs field attain masses of order $\langle S \rangle (M/m_P)^3$, we take their masses to be around the SUSY breaking scale $M_S$, which is fixed at $M_S = 10$ TeV in order to adequately suppress dimension five proton decay operator. The gauge coupling unification scale is set at $M_\text{GUT} \equiv (5/6)g_5 M = 10^{17}$ GeV, where $g_5$ is the unified gauge coupling of $SU(5)$.
\section{Summary}
We consider a pseudosmooth tribrid model of sneutrino inflation in an $R$-symmetric $SU(5)\times Z_5$ GUT model. With the help of an additional $Z_5$ symmetry and a non-minimal K\"{a}hler potential, a pseudosmooth trajectory is successfully generated to realize inflation while avoiding the monopole problem. The predicted values of the various inflationary parameters are calculated at the central value of the scalar spectral index, $n_s = 0.968$. The predictions for the tensor to scalar ratio, $ 2.7\times 10^{-3} \lesssim r  \lesssim 10^{-13}$, and for the running of the scalar spectral index, $-0.00031 \lesssim dn_s/d\ln k \lesssim 0.0024$, are in agreement with the latest Planck 2018 results. These ranges are obtained with $ -1\leq \kappa_h \leq -0.1$, $ 5.5 \times 10^{16} \lesssim M/\text{GeV} \lesssim  5.6 \times 10^{17}$, $ 5 \times 10^{16}\text{ GeV} \lesssim N_0 \lesssim  m_P$, $ |\gamma| \gtrsim 10^{-4}$, $|\delta| \lesssim 1$, $y_c = 1$ and $T_r = 10^6$ GeV. The gravitino problem is avoided with the realization of reheat temperature $T_r$ as low as $10^6$ GeV. A common problem of $R$-symmetric $SU(5)$ GUT is the appearance of light triplet and octet components from the GUT Higgs field, thus putting successful gauge coupling unification of MSSM in jeopardy. This problem is avoided with the help of additional vector-like families residing in complete multiplets,  ($5+\bar{5}+10+\bar{10}$), of $SU(5)$.

Before concluding we provide a few brief remarks related to proton decay and dark matter. Rapid proton decay from renormalizable superpotential couplings is not allowed in this $SU(5)$ model thanks to the $R$-symmetry. Furthermore, with relatively large squark and slepton masses of order $10$ TeV or so,  and with $M_\text{GUT} \sim 10^{17}$ GeV, dimension five proton decay is adequately suppressed, and dimension six proton decay mediated by the superheavy gauge bosons is predicted to lie well beyond the scope of Hyper-Kamiokande \cite{Abe:2011ts,Abe:2018uyc}. To realize the lightest supersymmetric particle (LSP) as a viable cold dark matter candidate, we need to separately invoke a $Z_2$ matter parity.
\section*{Acknowledgment}
This work is partially supported by the DOE grant No. DE-SC0013880 (Q.S.).

\end{document}